\documentclass[a4paper,envcountsame]{llncs}

\usepackage{todonotes}

\usepackage{afterpage}

\usepackage{array}
\makeatletter
\newcolumntype{"}{@{\hskip\tabcolsep\vrule width 1pt\hskip\tabcolsep}}
\makeatother

\usepackage{subcaption}
\usepackage{pgfplotstable}
\usepackage{graphicx}
\usepackage{amsfonts}
\usepackage{amsmath}
\usepackage{amssymb}
\usepackage{tikz}
\usepackage{xspace}
\usepackage{algorithm}
\usepackage{algpseudocode}
\usepackage{bimatrixgame}
\usepackage{gnuplottex}

\usepackage{rotating}
\usepackage{enumitem}

\newcommand{\eps}{\ensuremath{\epsilon}}

\newcommand{\PPAD}{\texttt{PPAD}\xspace}
\newcommand{\PSPACE}{\texttt{PSPACE}\xspace}

\newcommand{\timeout}{15 minutes\xspace}
\newcommand{\worstepsmix}{$0.3189$\xspace}
\newcommand{\worstepsts}{$0.3385$\xspace}
\newcommand{\worstepsksplus}{$0.4799$\xspace}
\newcommand{\worstndts}{$0.3254$\xspace}
\newcommand{\worstndport}{$0.3253$\xspace}

\newcommand{\blotnp}{Blotto-$n$-$p$\xspace}
\newcommand{\blotthreefive}{Blotto-3-5\xspace}
\newcommand{\blotthreeseven}{Blotto-3-7\xspace}
\newcommand{\blotthreenine}{Blotto-3-9\xspace}
\newcommand{\blotfourfive}{Blotto-4-5\xspace}
\newcommand{\blotfourseven}{Blotto-4-7\xspace}
\newcommand{\blotfournine}{Blotto-4-9\xspace}
\newcommand{\CovariantGamep}{CovariantGame-$p$\xspace}
\newcommand{\CovariantGameOne}{CovariantGame-1\xspace}
\newcommand{\CovariantGameThree}{CovariantGame-3\xspace}
\newcommand{\CovariantGameFive}{CovariantGame-5\xspace}
\newcommand{\CovariantGameSeven}{CovariantGame-7\xspace}
\newcommand{\CovariantGameNine}{CovariantGame-9\xspace}
\newcommand{\BertrandOligopoly}{BertrandOligopoly\xspace}
\newcommand{\BidirectionalLEG}{BidirectionalLEG\xspace}
\newcommand{\CournotDuopoly}{CournotDuopoly\xspace}
\newcommand{\CovariantGame}{CovariantGame\xspace}
\newcommand{\GrabTheDollar}{GrabTheDollar\xspace}
\newcommand{\GuessTwoThirdsAve} {GuessTwoThirdAve\xspace}
\newcommand{\LocationGame}{LocationGame\xspace}
\newcommand{\MinimumEffortGame}{MinimumEffortGame\xspace}
\newcommand{\RandomLEG}{RandomLEG\xspace}
\newcommand{\TravelersDilemma}{TravelersDilemma\xspace}
\newcommand{\RandomGame}{RandomGame\xspace}
\newcommand{\WarOfAttrition}{WarOfAttrition\xspace}
\newcommand{\Ranking}{Ranking\xspace}
\newcommand{\SGC}{SGC\xspace}
\newcommand{\Tournament}{Tournament\xspace}
\newcommand{\Unit}{Unit\xspace}

\newcommand{\LH}{LH\xspace}
\newcommand{\SE}{SE\xspace}
\newcommand{\Pure}{Pure\xspace}
\newcommand{\DMP}{DMP\xspace}
\newcommand{\BBMone}{BBM1\xspace}
\newcommand{\BBMtwo}{BBM2\xspace}
\newcommand{\TStwo}{TS2\xspace}
\newcommand{\TSone}{TS001\xspace}
\newcommand{\KS}{KS\xspace}
\newcommand{\KSplus}{KS+\xspace}

\title{An empirical study of finding approximate equilibria in bimatrix games%
}

\author{John Fearnley
\and
Tobenna Peter Igwe
\and
Rahul Savani}

\institute{Department of Computer Science, University of Liverpool, UK.}

\begin{document}

\maketitle


\begin{abstract}
While there have been a number of studies about the efficacy of methods to find exact Nash equilibria in bimatrix games, there has been little empirical work on finding approximate Nash equilibria. Here we provide such a study that compares a number of approximation methods and exact methods. In particular, we explore the trade-off between the quality of approximate equilibrium and the required running time to find one. We found that the existing library GAMUT, which has been the de facto standard that has been used to test exact methods, is insufficient as a test bed for approximation methods since many of its games have pure equilibria or other easy-to-find good approximate equilibria. We extend the breadth and depth of our study by including new interesting families of bimatrix games, and studying bimatrix games upto size $2000 \times 2000$.  Finally, we provide new close-to-worst-case examples for the best-performing algorithms for finding approximate Nash equilibria. 
\end{abstract}

\section{Introduction}

The computation of Nash equilibria in bimatrix games is one of the central
topics in game theory, which has received much attention 
from a theoretical point of view.  It has been shown that the
problem of finding a Nash equilibrium is \PPAD-complete~\cite{DGP,CDT}, which
implies that we are unlikely to find a polynomial-time algorithm for this
problem. Naturally, this has led to a line of work studying the complexity of
finding \emph{approximate} Nash equilibria~\cite{DMP,Bosse,TS,KS,fearnley12}. 

Most of the work on approximation algorithms has focussed on additive
approximations, where two different notions are used. An \emph{$\epsilon$-Nash
equilibrium} requires that both players achieve an expected payoff that is
within $\epsilon$ of a best response, while the stronger notion of
\emph{$\epsilon$-well supported Nash equilibrium} ($\epsilon$-WSNE) requires
that both players only place probability on strategies that are within
$\epsilon$ of a best response. The current state of the art for $\epsilon$-Nash
equilibria is the algorithm of Tsaknakis and Spirakis~\cite{TS}, which finds a
$0.3393$-Nash equilibrium in polynomial time, and the current state of the art
of $\epsilon$-WSNE is the algorithm of
Fearnley et al.~\cite{fearnley12}, which finds a
$(\frac{2}{3} - 0.00591)$-WSNE in polynomial time.

So far, most of the work on approximate equilibria has been theoretical in
nature. The goal of this paper is to answer the following question: \emph{
Are approximate equilibria relevant to the problem of solving bimatrix
games in practice?} To answer this, we must study several related questions.
\begin{itemize}[leftmargin=0pt]
\item Firstly, how good are the algorithms for finding exact Nash equilibria in
practice?  If they are good enough, then there is no need for approximation.
Otherwise, how much faster are the approximation algorithms? 
\item Secondly, what
quality of approximation do the approximation algorithms provide in practice?
If the best theoretical guarantee of a $0.3393$-Nash equilibrium is not beaten
in practice, it is unlikely to be useful.
\item Finally, is there a trade off between running time and
approximation? We have a wide variety of approximation algorithms, from those
that solve a single linear program, to those that perform complicated gradient
descent procedures. Do fast algorithms generally produce worse approximate
equilibria? Should our desired quality of equilibrium impact our choice of
algorithm?
\end{itemize}

\noindent
\textit{Our contribution.}
While there have been several empirical studies on computing exact
equilibria~\cite{CdRP08,GPRS12,PNS08,SGC05}, the empirical work for
approximate equilibria has so far been limited to a paper~\cite{TSK08}
that evaluates the algorithm of Tsaknakis and Spirakis (TS)~\cite{TS},
and one that looks exclusively at symmetric games~\cite{KS11}. We
address this by performing a comprehensive study of approximation algorithms. We
compare the performance of five different algorithms for finding approximate
equilibria on 15 different types of game. Moreover, we include two algorithms
for finding exact equilibria: the Lemke-Howson algorithm and support
enumeration.

This allows us to answer the questions posed earlier.
Firstly, we find that approximation algorithms can tackle 
instances that exact algorithms cannot. With a timeout of
\timeout, we found that exact algorithms were mostly unable to solve instances
of size $1000 \times 1000$, whereas approximation algorithms could
easily tackle instances of size $2000 \times 2000$.
Secondly, we find that approximation algorithms 
often perform much better than their theoretical worst case guarantees: in
agreement with the experimental study of Tsaknakis et al. \cite{TSK08}, we find
that the TS algorithm often finds $0.01$-Nash equilibria or better. In answer to
the third question: while our data shows that the TS algorithm clearly wins in
terms of quality of approximation, it is usually the slowest, and if we only
require weaker approximate equilibria, then other algorithms can find one
faster. 

To obtain our results, we tested the algorithms on a wide variety of
games. Previous work on exact equilibria has typically used the GAMUT
library~\cite{NWSL-B04}. However, almost all of the games provided by GAMUT have
exact pure Nash equilibria, so using only these games could skew our results.
For example, the work of Porter et al.~\cite{PNS08} concluded that support
enumeration typically outperforms the Lemke-Howson algorithm, based on
the fact that support enumeration can quickly find the pure equilibria in the
games provided by GAMUT.

There are many practical applications of game theory where all equilibria use
mixed strategy profiles. Our second contribution is to define several natural
classes of games that do not have pure Nash equilibria. 
All of our game generators and algorithm implementations are open source and
freely available\footnote{\texttt{http://bimatrix-games.github.io/}}. Our
results show that algorithms perform very differently on these games, so they should be included in any future study of the practical aspects of
computing equilibria.

The excellent performance of the TS algorithm raises the question 
of whether the upper bound of 0.3393 on the quality of approximation of this 
algorithm is tight. We used a genetic algorithm to
search for worst-case examples, and we found a $5\times 5$ bimatrix
game in which the TS algorithm gives a \worstepsts-Nash
equilibrium, which shows that the performance guarantee is essentially tight.
We applied this technique to the algorithm of
Fearnley et al.~\cite{fearnley12} for finding
$\epsilon$-WSNE, which likewise had no good theoretical
lower bound. However, we were only able to find an example for which the
algorithm gives a \worstepsksplus-WSNE, which is far from the 
upper bound of $(\frac{2}{3} - 0.00591375)$. 
Finally, to test the limits of approximation techniques, we ran the
same procedure against the combination of our three best approximation
algorithms (the TS algorithm, the algorithm of Bosse et al. \cite{Bosse}, and
the best pure strategy pair). Here the fitness
function was the minimum of the approximations provided by the algorithms. We
found a game for which all of the algorithms gave no better
than a \worstepsmix-Nash equilibrium, which is relatively close to the
theoretical upper bound of $0.3393$, and indicates that 
new techniques will be needed to advance the theoretical state of the art.

\vskip \baselineskip

\textit{Related work.}
For exact equilibria, Porter et al.~\cite{PNS08} use GAMUT to
compare support
enumeration (SE) with the Lemke-Howson (LH) algorithm. They showed that SE
performs well when compared to LH, because many of the games in the library have
small support equilibria. They also highlight random games and covariant games
as the most challenging GAMUT games for SE and LH, which our results also
confirm. Most of their experiments considered games of size $600 \times 600$,
but they did consider random games at game sizes up to $1000 \times 1000$.

Sandholm at al.~\cite{SGC05} describe four ways of solving games via mixed
integer programming (MIP). Experiments were carried out games of size
$150 \times 150$ provided by GAMUT. It was found that MIP
performed better than LH but was outclassed by SE.

Codenotti et al.~\cite{CdRP08} studied of LH algorithm,
on random and covariant games, where it was found
that LH has a running time of $O(n^7)$
for $n\times n$ covariant games. They presented a heuristic which
involves running different LH paths in parallel,
which showed an 
improvement over LH for random games, but not
covariant games. They looked at games
of size up to $1000 \times 1000$.

Tsaknakis et al.~\cite{TSK08} performed an experimental analysis of their
algorithm (TS)
for finding a $0.3393$-Nash equilibrium in polynomial time.
They studied games of size $100 \times 100$, and they also
constructed games with no small support equilibria, to prevent easy
solutions. 
It was found that TS always finds a $0.015$-Nash
equilibrium or better. We confirm the result that in
general TS performs well, however among some of our games TS only finds a
$0.14$-Nash equilibrium. 

Gatti et al.~\cite{GPRS12} evaluated the performance of LH, MIP and SE.  They
introduced a number of heuristics, and compared their performance on the games
from GAMUT. They found that none of the methods was superior for all games. They
did look at the quality of approximation achieved by their heuristics and
algorithms, but the largest games they looked at were of size $150 \times 150$.


\section{Experimental Setup}

\noindent
\textit{Algorithms.}
Now, we describe the algorithms that are studied in this paper. For
the computation of exact equilibria, we study the following two algorithms.

\begin{itemize}[leftmargin=0pt]

\item \textbf{\LH.} The Lemke-Howson algorithm is a widely-used pivoting
algorithm~\cite{lemke64}. 
It has
exponential worst case behaviour~\cite{savani06}, and it is 
\PSPACE-hard to compute the equilibrium that it finds~\cite{GPS13}. We use the
implementation from \cite{CdRP08}, modified so that degeneracy is resolved by the lexicographic minimum ratio test.


\item \textbf{\SE.} Support enumeration is brute force algorithm that,
 for every possible pair of supports, solves a system of linear equations
to check for an equilibrium. Our implementation goes through
supports from small to large cardinality.
	
\end{itemize}

There is a wide variety of algorithms for finding $\eps$-Nash equilibria,
and we have implemented the following (polynomial-time) algorithms:

\begin{itemize}[leftmargin=0pt]

\item \textbf{\Pure.} This algorithm checks all pure strategy profiles and
	returns one that gives the best $\epsilon$-Nash equilibrium. 

\item \textbf{\DMP.} This algorithm was given by Daskalakis et al.~\cite{DMP}.
It finds a $0.5$-Nash equilibrium using a very simple approach that starts with
an arbitrary pure strategy, and then makes two best response queries. 


\item \textbf{\BBMone.} This is the first of two algorithms given by Bosse et
	al.~\cite{Bosse}. It finds a $0.3819$-Nash equilibrium. Given a bimatrix
	game $(R, C)$, \BBMone solves the zero-sum game $(R-C, C-R)$ using
	linear programming. Then it proceeds in a similar manner to \DMP, but uses
the LP solution as the initial strategy.

\item \textbf{\BBMtwo.} This is the second of the two algorithms given by Bosse et
	al.~\cite{Bosse}. It finds a $0.3639$-Nash equilibrium. It is 
	an adaptation of \BBMone that contains some extra steps to deal
	with cases where the first algorithm performs poorly.

\item \textbf{TS.} This algorithm was given by Tsaknakis and Spirakis~\cite{TS}.
	It finds a $(0.3393+\delta)$-Nash equilibrium, where $\delta$ is an
	arbitrary positive constant. The algorithm uses gradient descent over
	the space of mixed strategy profiles. The objective function is the quality
of approximate Nash equilibrium. The algorithm finds a stationary point. If
	the stationary point is not a $(0.3393+\delta)$-Nash equilibrium, then it can be used to find a second point that is a
	$(0.3393+\delta)$-Nash equilibrium. To investigate the dependence of the
	algorithm on $\delta$, we use two versions with $\delta = 0.2$ and $\delta =
	0.001$. We refer to these as \textbf{\TStwo} and \textbf{\TSone},
	respectively. At the end of Section~\ref{sec:exp_results}, we analyze the effect of the choice of
	$\delta$.
	
\end{itemize}

There has been comparatively less study of algorithms to find approximate
well-supported Nash equilibria. We implemented the following two algorithms:


\begin{itemize}[leftmargin=0pt]

\item \textbf{\KS.} This algorithm was given by Kontogiannis and Spirakis, and it
	finds a $\frac{2}{3}$-WSNE~\cite{KS}. The algorithm first checks all pure
	strategy profiles in order to determine if there is a pure
	$\frac{2}{3}$-WSNE. If not, then the algorithm solves the same zero-sum game
	as \BBMone/\BBMtwo, and the equilibrium of this game is a $\frac{2}{3}$-WSNE.

\item \textbf{\KSplus.} This algorithm gives an improved approximation guarantee
	compared to \KS~\cite{fearnley12} of $(\frac{2}{3} -
	0.00591)$. It combines the \KS algorithm with two extra
	procedures: one that finds the best WSNE with $2 \times 2$
	support, and one that finds the best WSNE on the supports from an equilibrium 
	to the KS zero-sum game. 

\end{itemize}

\paragraph{Game classes.}
We now describe the classes of games used in our study.
In order to have a consistent meaning of approximation guarantees, all
games are scaled to have payoffs in $[0,1]$.
Firstly, we used games provided by the GAMUT library. We used every class
of games in GAMUT that could be scaled indefinitely. We eliminated the
classes of games that have fixed size, and we were also forced to eliminate some
classes of games because their generators either crashed or produced invalid
games when asked to produce games with more than 1000 strategies per
player, which included \BidirectionalLEG and \RandomLEG.
We were left with the games shown below.

\begin{center}
\begin{tabular}{l@{\qquad}l@{\qquad}l}
\textbf{\BertrandOligopoly} & \textbf{\CournotDuopoly}  &
\textbf{\CovariantGame} \\
\textbf{\GrabTheDollar} & \textbf{\GuessTwoThirdsAve} & \textbf{\MinimumEffortGame} \\
\textbf{\TravelersDilemma} & \textbf{\RandomGame}  & \textbf{\WarOfAttrition}
\end{tabular}
\end{center}

In covariant games each pure strategy profile is drawn from a multivariate
normal distribution with covariance $\rho$.  When
$\rho = 1$ we have a coordination game and when $\rho = -1$ we have a zero-sum
game. Previous work~\cite{PNS08,CdRP08} indicates that these games are easy to
solve when $\rho > 0$, with the hardest games in the range $[-0.9, -0.5]$.
We study 5 classes of
games \CovariantGamep for $p=1,3,5,7,9$ where $\rho=-0.1,-0.3,-0.5,-0.7,-0.9$
respectively.

With the exception of covariant and random games, the other bimatrix game
classes provided by GAMUT have pure equilibria, and are therefore easily solved
by support enumeration. In order to broaden our study, we chose to implement
generators for the following games, which generally do not have pure equilibria.

\begin{itemize}[leftmargin=0pt]

\item \textbf{Non-zero sum colonel \textbf{Blotto} Games.} The players have an
equal number of soldiers $T$ that must be assigned simultaneously to $n$
hills. Each player has a value for each hill that he
receives if he assigns strictly more soldiers to the hill than
his opponent (ties are broken uniformly at random.)
Each player's payoff is the sum of the value of the hills
won by that player. To avoid pure equilibria, the hill values are drawn from a
multivariate normal distribution with covariance of $\rho > 0$. In our
experiments, we study families of games with $n=3,4$ and $\rho= 0.5, 0.7, 0.9$,
which we denote by \blotnp for $p=5,7,9$, respectively.  The number of soldiers
$T$ was varied in order to generate a scalable family of games. 

\item \textbf{\textbf{Ranking} Games}~\cite{GGKV13}. Each
	player chooses an effort level with an associated cost and score. A prize is
given to the player with the higher score, or is split in the case of a tie.
The payoff of a player is the value of the prize minus the cost of the
effort. We generated scores and costs as increasing step
functions of effort with random step sizes. We denote these games by \Ranking.

\item \textbf{SGC games.} Sandholm et al.~\cite{SGC05} also noted that most
GAMUT games have small support equilibria. They
introduced a family of games where, in all equilibria, both players use
half of their actions. We denote their games as \SGC. In
these game the only equilibrium in a $(2k - 1) \times (k - 1)$  game has support
sizes~$k$ for both players, which makes these games hard for support
enumeration.

\item \textbf{Tournament Games}~\cite{Anbalaga13}. 
	Starting with a random tournament, an asymmetric bipartite
	graph is constructed 
where one side corresponds to the nodes of the tournament, and the other
corresponds to subsets of nodes.
The bipartite graph is transformed into a win-lose
	game where the actions of each player are the nodes on their
	side of the graph. We denote these games
	as \Tournament.

\item \textbf{Unit Vector Games (UVG)}~\cite{SvS15}. The payoffs for the
	column player are chosen randomly from the range $[0, 1]$, but for the row
player
each column $j$ contains exactly one $1$ payoff
	with the rest being $0$. In order to avoid pure equilibria, we generated
	these games by placing the $1$s uniformly at random in the rows that do not
	generate a pure equilibrium. We denote these games as \Unit.

\end{itemize}

Some of these games are not square. So, in our results we use instances that
have roughly the same name number of payoffs as the corresponding square games,
e.g., we compare $100 \times 100$ games to non-square games with roughly $10000$
payoff entries. 

Unlike other studies~\cite{GPRS12,SGC05}, we do not include the exponential-time
examples for \LH devised by
Savani and von Stengel~\cite{savani06}. They are not suitable for
an experimental study on approximate equilibria because the games have a number
of large
payoffs, so when they are normalised to the range $[0, 1]$, almost all payoffs
in the game are close to $0$. Hence these games are very easy to approximate.
For example, 
the $30 \times 30$ instance has a pure $1.63 \times 10^{-10}$-WSNE. Furthermore,
for instances
larger than $16 \times 16$, these games exhaust the precision of floating point.


\vskip 0.5\baselineskip
\noindent
\textit{Implementation details.}
All implementations are written in C. We used CPLEX 
to solve the linear programs used in some of the
algorithms. For our runtime results, we only measured the amount of time spent
by the solver, and discarded the time taken to read the game from its input
file. Our experiments were carried out on a cluster of 8 identical machines
running Scientific Linux 6.6, which each have an Intel Core i5-2500k processor
clocked at 3.30GHz with 16GB of RAM.

To verify our results, we implemented three programs that compute the quality of
exact, approximate, and well-supported Nash equilibria, respectively. All of
these programs carry out their calculations in \emph{exact arithmetic}. Our
exact equilibrium checker takes a pair of supports, and checks whether there
exists a Nash equilibrium on these supports. Both of the approximate checkers
take a mixed strategy profile, and output the value of $\epsilon$ that this
profile achieves.

\section{Experimental Results}
\label{sec:exp_results}



\paragraph{Exact Algorithms.}
We tested \LH and \SE against our library of games with a timeout of \timeout.
Table~\ref{tbl:main} part A shows the percentage of games that were solved by
these two algorithms for various game sizes. We
have divided the games into three classes. Firstly we have 
GAMUT games that always have pure equilibria. As expected, \SE performs
well on these games, while \LH is also able to tackle the majority of
instances. Secondly, we have the GAMUT games that do not always have pure
equilibria. Both algorithms performed very poorly for
covariant games, which is in agreement with previous studies. Finally, we have
the games that we proposed. Both algorithms
struggle with the games in this class, which supports the idea
that GAMUT's existing library does not give a comprehensive picture of possible
games. \Ranking games provide an interesting case that differentiates between
\LH and \SE: these games only had equilibria with medium sized support so \SE
was hopeless, however \LH was able to solve these games using a linear number of
pivots. In conclusion, our results show that exact methods are
inadequate for the 2nd and 3rd classes of games, so it is for these games that
we are interested in the performance of approximation methods.



\renewcommand{\arraystretch}{1}
\begin{sidewaystable}
{\small
\begin{tabular}{cccc}
& A: \% completed within 15mins & B: $\eps$-Nash & C: $\eps$-WSNE \\
\begin{tabular}{|r|}
  \hline
 \\ 
  \hline
 \\
\BertrandOligopoly \\
  \CournotDuopoly \\
  \GrabTheDollar \\
  \GuessTwoThirdsAve \\
  \LocationGame \\
  \MinimumEffortGame \\
  \TravelersDilemma \\
  \WarOfAttrition \\
\hline
  \CovariantGameOne \\
  \CovariantGameThree \\
  \CovariantGameFive \\
  \CovariantGameSeven \\
  \CovariantGameNine \\
  \RandomGame \\
\hline
  \blotthreefive \\
  \blotthreeseven \\
  \blotthreenine \\
  \blotfourfive \\
  \blotfourseven \\
  \blotfournine \\
  \SGC \\
  \Ranking \\
  \Tournament \\
  \Unit \\
   \hline
\end{tabular}
 &
\begin{tabular}{|r|r|r"r|r|r|}
  \hline
 \multicolumn{3}{|c"}{LH} & \multicolumn{3}{c|}{SE}\\
\hline
 100 & 300 & 1000 &100 & 300 & 1000\\
  \hline
80 & 35 & 12 & 96 & 100 & 96 \\ 
  100 & 100 & 68 & 100 & 100 & 68 \\ 
  100 & 100 & 100 & 100 & 100 & 100 \\ 
  36 & 20 & 0 & 100 & 100 & 100 \\ 
  100 & 100 & 100 & 100 & 100 & 96 \\ 
  100 & 100 & 100 & 100 & 100 & 100 \\ 
  100 & 75 & 56 & 100 & 100 & 100 \\ 
  100 & 100 & 100 & 100 & 100 & 100 \\ 
\hline
  100 & 50 & 4 & 82 & 62 & 24 \\ 
  100 & 13 & 0 & 10 & 25 & 0 \\ 
  100 & 0 & 0 & 0 & 0 & 0 \\ 
  100 & 0 & 0 & 0 & 0 & 0 \\ 
  100 & 19 & 0 & 0 & 0 & 0 \\ 
  100 & 73 & 17 & 88 & 80 & 56 \\ 
\hline
  80 & 48 & 48 & 56 & 60 & 72 \\ 
  68 & 48 & 60 & 48 & 56 & 56 \\ 
  52 & 40 & 52 & 56 & 48 & 56 \\ 
  76 & 44 & 32 & 40 & 24 & 40 \\ 
  68 & 44 & 44 & 32 & 12 & 32 \\ 
  76 & 52 & 36 & 8 & 12 & 16 \\ 
  100 & 100 & 100 & 0 & 0 & 0 \\ 
  100 & 100 & 100 & 0 & 0 & 0 \\ 
  16 & 12 & 12 & 16 & 0 & 0 \\ 
  100 & 96 & 44 & 0 & 0 & 0 \\ 
   \hline
\end{tabular}
 &
\begin{tabular}{|r"r|r"r|r"r|r"r|r"r|r"r|}
  \hline
  \multicolumn{2}{|c|}{DMP} & \multicolumn{2}{c|}{BBM1} & \multicolumn{2}{c|}{BBM2} & \multicolumn{2}{c|}{Pure} & \multicolumn{2}{c|}{TS001} & \multicolumn{2}{c|}{TS2} \\ 
  \hline
  Time & $\eps$ & Time & $\eps$ & Time & $\eps$ & Time & $\eps$ & Time & $\eps$ & Time & $\eps$ \\
  \hline
   0.00 & 0.42 & 0.43 & 0.00 & 0.47 & 0.00 & 0.04 & 0.00 & 318.34 & 0.00 & 20.96 & 0.06 \\ 
   0.00 & 0.25 & 0.42 & 0.00 & 0.43 & 0.00 & 28.14 & 0.00 & 1.95 & 0.00 & 2.94 & 0.00 \\ 
   0.00 & 0.25 & 0.77 & 0.11 & 0.78 & 0.10 & 0.00 & 0.00 & 3.31 & 0.00 & 5.08 & 0.00 \\ 
   0.00 & 0.50 & 0.70 & 0.00 & 0.71 & 0.00 & 0.00 & 0.00 & 2.55 & 0.00 & 2.94 & 0.00 \\ 
   0.00 & 0.20 & 0.76 & 0.00 & 0.78 & 0.00 & 147.13 & 0.00 & 3.22 & 0.00 & 4.87 & 0.00 \\ 
   0.00 & 0.00 & 0.78 & 0.00 & 0.78 & 0.00 & 0.00 & 0.00 & 30.76 & 0.00 & 49.03 & 0.00 \\ 
   0.00 & 0.16 & 0.72 & 0.00 & 0.78 & 0.00 & 0.00 & 0.00 & 233.38 & 0.00 & 134.10 & 0.00 \\ 
   0.00 & 0.00 & 0.99 & 0.00 & 1.04 & 0.00 & 0.00 & 0.00 & 4.13 & 0.00 & 4.30 & 0.00 \\ 
\hline
   0.00 & 0.21 & 59.04 & 0.01 & 57.43 & 0.01 & 146.50 & 0.02 & 125.35 & 0.00 & 572.97 & 0.00 \\ 
   0.00 & 0.22 & 56.75 & 0.01 & 58.08 & 0.01 & 150.07 & 0.04 & 130.26 & 0.00 & 671.48 & 0.00 \\ 
   0.00 & 0.27 & 58.76 & 0.01 & 61.78 & 0.01 & 149.42 & 0.11 & 235.68 & 0.00 & 727.18 & 0.00 \\ 
   0.00 & 0.29 & 56.62 & 0.01 & 53.85 & 0.01 & 148.41 & 0.14 & 330.66 & 0.00 & 768.42 & 0.00 \\ 
   0.00 & 0.32 & 54.88 & 0.00 & 57.90 & 0.00 & 153.06 & 0.21 & 529.19 & 0.00 & 743.63 & 0.00 \\ 
   0.00 & 0.35 & 56.38 & 0.03 & 59.06 & 0.03 & 89.94 & 0.00 & 228.91 & 0.00 & 550.99 & 0.00 \\ 
\hline
   0.00 & 0.30 & 6.76 & 0.10 & 6.86 & 0.10 & 67.63 & 0.06 & 18.89 & 0.00 & 10.29 & 0.03 \\ 
   0.00 & 0.37 & 5.75 & 0.11 & 5.95 & 0.11 & 74.69 & 0.10 & 22.92 & 0.00 & 10.30 & 0.04 \\ 
   0.00 & 0.38 & 1.24 & 0.05 & 1.23 & 0.05 & 100.08 & 0.12 & 17.19 & 0.00 & 11.49 & 0.02 \\ 
   0.00 & 0.38 & 1.41 & 0.13 & 1.39 & 0.13 & 117.29 & 0.10 & 24.33 & 0.00 & 9.28 & 0.04 \\ 
   0.00 & 0.38 & 2.18 & 0.12 & 2.10 & 0.13 & 118.91 & 0.13 & 28.19 & 0.00 & 12.75 & 0.05 \\ 
   0.00 & 0.40 & 2.12 & 0.10 & 2.07 & 0.10 & 164.55 & 0.18 & 19.89 & 0.00 & 13.25 & 0.03 \\ 
   0.00 & 0.25 & 3.46 & 0.00 & 2.87 & 0.00 & 121.55 & 0.25 & 57.02 & 0.00 & 63.53 & 0.00 \\ 
   0.00 & 0.32 & 6.82 & 0.15 & 7.00 & 0.15 & 148.89 & 0.18 & 207.00 & 0.00 & 45.55 & 0.06 \\ 
   0.23 & 0.50 & 1.95 & 0.06 & 1.96 & 0.06 & 374.25 & 1.00 & 39.37 & 0.00 & 61.92 & 0.01 \\ 
   0.00 & 0.39 & 57.70 & 0.01 & 57.77 & 0.01 & 153.08 & 0.00 & 46.74 & 0.00 & 32.96 & 0.00 \\ 

   \hline
\end{tabular}
 &
\begin{tabular}{|r"r|}
  \hline
\multicolumn{2}{|c|}{KS}\\
  \hline
Time & $\eps$ \\ 
  \hline
0.04 & 0.00 \\ 
  28.14 & 0.00 \\ 
  0.00 & 0.00 \\ 
  0.00 & 0.00 \\ 
  147.13 & 0.00 \\ 
  0.00 & 0.00 \\ 
  0.00 & 0.00 \\ 
  0.00 & 0.00 \\ 
\hline
  198.30 & 0.01 \\ 
  208.96 & 0.01 \\ 
  208.35 & 0.01 \\ 
  203.86 & 0.01 \\ 
  215.01 & 0.01 \\ 
  109.11 & 0.00 \\ 
\hline
  67.50 & 0.03 \\ 
  75.22 & 0.04 \\ 
  100.56 & 0.03 \\ 
  120.78 & 0.10 \\ 
  123.28 & 0.10 \\ 
  167.79 & 0.11 \\ 
  119.16 & 0.00 \\ 
  158.68 & 0.18 \\ 
  388.14 & 0.07 \\ 
  211.65 & 0.00 \\ 
   \hline
\end{tabular}
 
\end{tabular}
}
\caption{
A shows the percentage of instances which did not
time out on LH and SE for instances of various sizes (we used sizes 105, 300 1035 for Blotto-3 games, 120, 364 and 969
for Blotto-4 games, 27, 57, 126 rows for \Tournament and 100, 300, 1000 for all
other games.)
B and C show experiment results on games of size $2000 \times 2000$ (games of
sizes 2016 for Blotto-3, 2024 for Blotto-4, 1999 for SGC, 200 rows for
Tournament.)
We display the average running time (in seconds) and average 
$\epsilon$-NE ($\epsilon$-WSNE for KS) across instances which did not time out (of the approximation
algorithms only \TSone ever timed out - 17\% of the time). 
}
\label{tbl:main}
\end{sidewaystable}

\vskip 0.5\baselineskip
\textit{Approximation algorithms.}
Table~\ref{tbl:main} part B shows the running time and quality of
approximation, respectively, when the algorithms were tested on games of size
$2000 \times 2000$.  Again, there is a clear split in the data between the
``easy'' games from GAMUT and the more challenging game classes. Note that, with
only a handful of exceptions, the approximation algorithms were easily able to
deal with games of this size. Only \TSone was observed to time out, and it timed
out on 17\% of the instances that it was tested on. This indicates that
approximation algorithms can indeed be applied to games that cannot be tackled
with exact methods. We summarize the performance of the algorithms as
follows.

\begin{itemize}[leftmargin=0pt]

\item \textbf{\DMP.} This algorithm runs
	quickly and usually gives a poor approximation. In terms of
	quality, it is clearly outmatched by all of the other algorithms.

\item \textbf{\BBMone and \BBMtwo.} These algorithms were typically in the
	middle in terms of both approximation quality and running time. Only a
	handful of Blotto instances triggered the extra steps in \BBMtwo,
	so these two algorithm are mostly identical.

\item \textbf{\Pure.} Whenever the game has a pure NE, this
	algorithm performs well, because it terminates once a pure NE has been
	found.  Otherwise, it is among the slowest of the algorithms,
	because it is never faster than $n^3$, where $n$ is the number
of strategies. The quality of
	approximation results confirm that our new games succeed in avoiding
	pure strategy profiles that are close to being Nash equilibria.

\item \textbf{\TStwo and \TSone.} The TS algorithm was the clear winner in terms
	of quality of approximation. The results show that the choice of $\delta$
	can have a significant effect on the algorithm's characteristics. \TStwo
	often terminates in a reasonable running time when compared to BBM, and it
	usually beats BBM significantly on quality of approximation. However, \TSone
	always beats \TStwo in quality of approximation, and always provides the
	best approximations among all of the algorithms that we studied. This
	accuracy comes at the cost of speed, as there are many games upon which
	\TSone is slower than \TStwo.

\end{itemize}

Table~\ref{tbl:main} shows the results for the WSNE algorithms on the
same set of games. Due to its $O(n^4)$ running time, we found that
that \KSplus timed out on all instances, so 
results for this algorithm are omitted.
Recall that \KS uses a preprocessing step to search for pure
WSNE, and then solves an LP.
We found that the preprocessing almost always provides the better
approximation, but the search over pure strategies is the dominant
component of \KS's running time. So, there is a 
significant cost for targeting $\epsilon$-WSNE over $\epsilon$-Nash equilibria,
since \Pure is the slowest algorithm for $\epsilon$-Nash equilibria,
and \KS is never faster than \Pure.

\begin{figure}[ht]
\includegraphics[page=2,width=\textwidth]{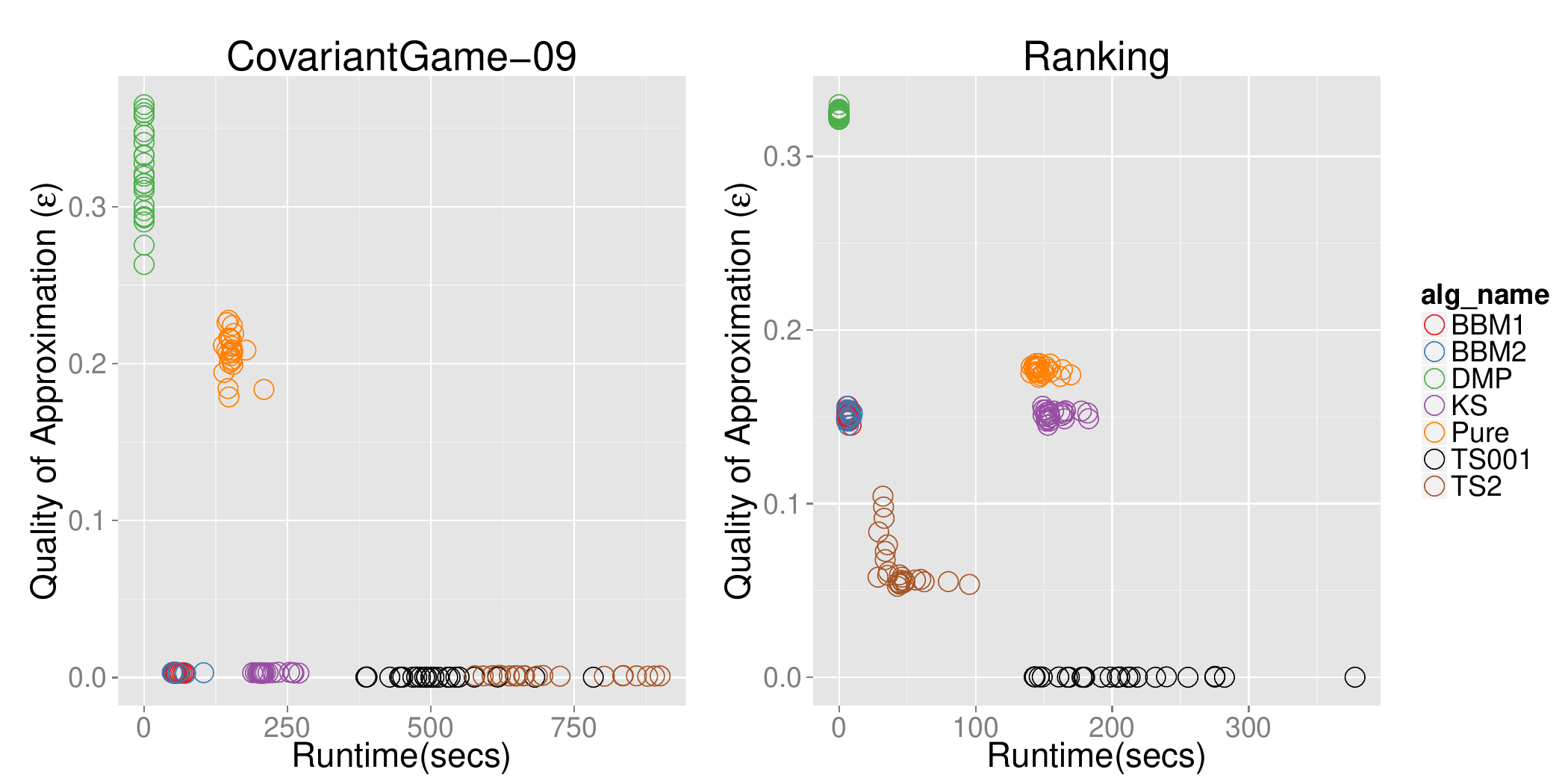}
\caption{Runtime vs Quality of approximation for \CovariantGameNine and \Ranking 
for instances of size $2000 \times 2000$.
}
\label{fig:epsrunning}
\end{figure}

Finally, we comment on the trade off between running time and quality of
approximation. In Figure~\ref{fig:epsrunning} we plot these two metrics against
each other for \CovariantGameNine and \Ranking, which provide a fairly
representative sample of the results that we observed across the dataset. The
points towards the lower left of the diagrams are those that minimize the
running time for a given approximation guarantee. In order of accuracy, we
typically see points from \DMP, \BBMone, \TStwo, and \TSone along this frontier.

\begin{figure}[ht]
\begin{center}
\includegraphics[scale=0.7]{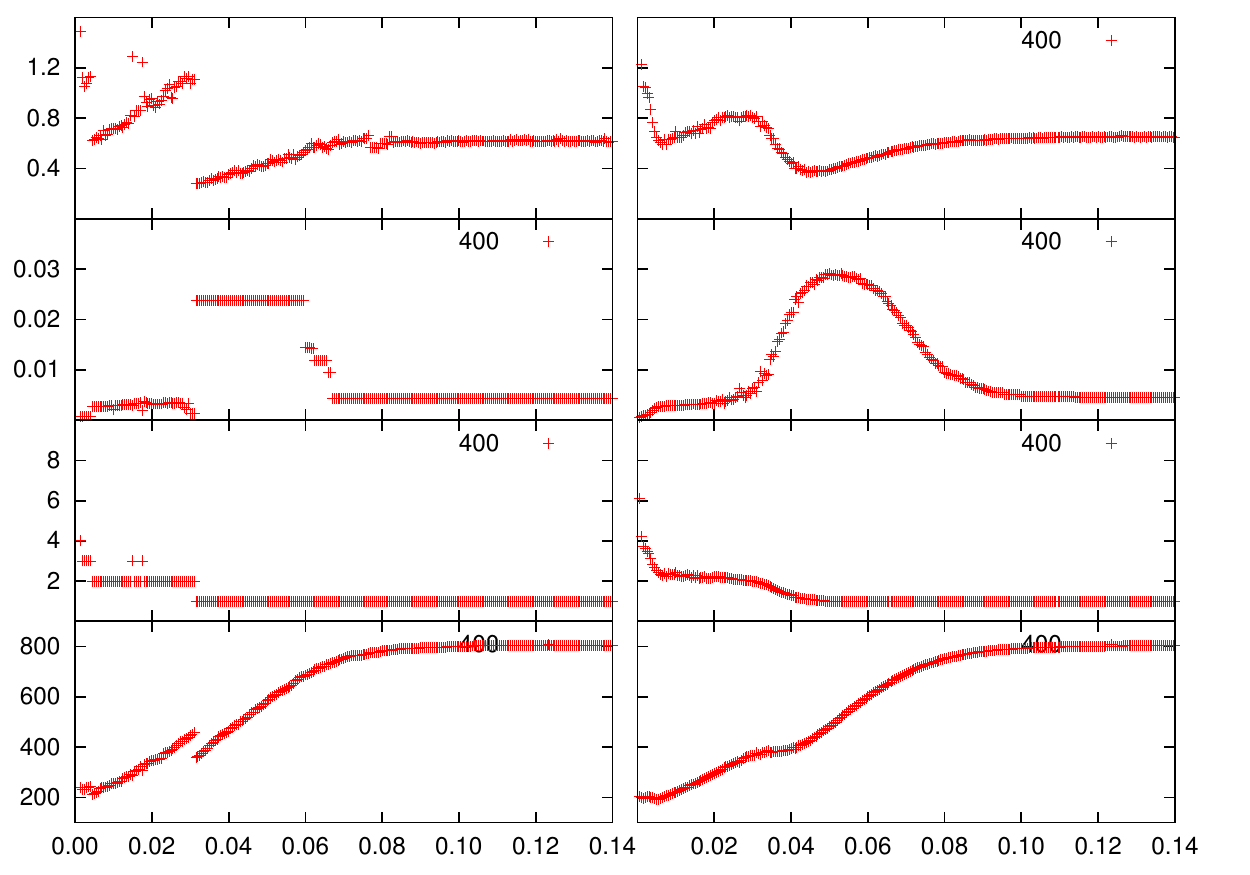}
\end{center}
\caption{TS performance plots of runtime (row 1), quality of approximation
(row 2), number of iterations (row 3,) and
average LP-Size (row 4) against $\delta$. Left diagrams shows
results for a single $400 \times 400$ instance of \RandomGame, while right diagrams show averages
over one hundred $400 \times 400$ instances of \RandomGame.}
\label{fig:ts_delta}
\end{figure}

\vskip 0.5\baselineskip
\noindent
\textit{The TS algorithm.}
Since our results indicate that the TS algorithm gives the best 
approximations, it is worth spending more time analysing this algorithm. Both
the quality of approximation and the running time of the
algorithm are affected by the choice of $\delta$. We now give more detailed
results on how this parameter affects the of the algorithm. To
test the dependence on $\delta$, we ran the TS algorithm on one hundred $400
\times 400$ instances of \RandomGame for various values of $\delta$ in the range
$(0, 0.14]$. The results of these experiments are displayed in
Figure~\ref{fig:ts_delta}. The left side of the figure shows the results for a
single game, while the right side of the figure shows the average results over
all instances.

The first two rows show the runtime and quality of approximation, respectively.
It can be seen that the algorithm does not scale smoothly with respect to
$\delta$, and instead there are discontinuities in both running time and quality
of approximation. The explanation for these discontinuities can be found in the
third and fourth rows, which show the number of rows and the size of the LP that
is solved in each iteration, respectively. The third row shows that, as we would
expect, the number of iterations increases as $\delta$ decreases. However, the
data in the fourth row shows that the story is more complicated. 
The size of the LP that is formulated in each iteration increases as $\delta$
increases. Thus, although the number of iterations falls, the time per iteration
gets larger. 

\section{Worst case examples}

\begin{figure}[ht]
\centering
\includegraphics[scale=0.50]{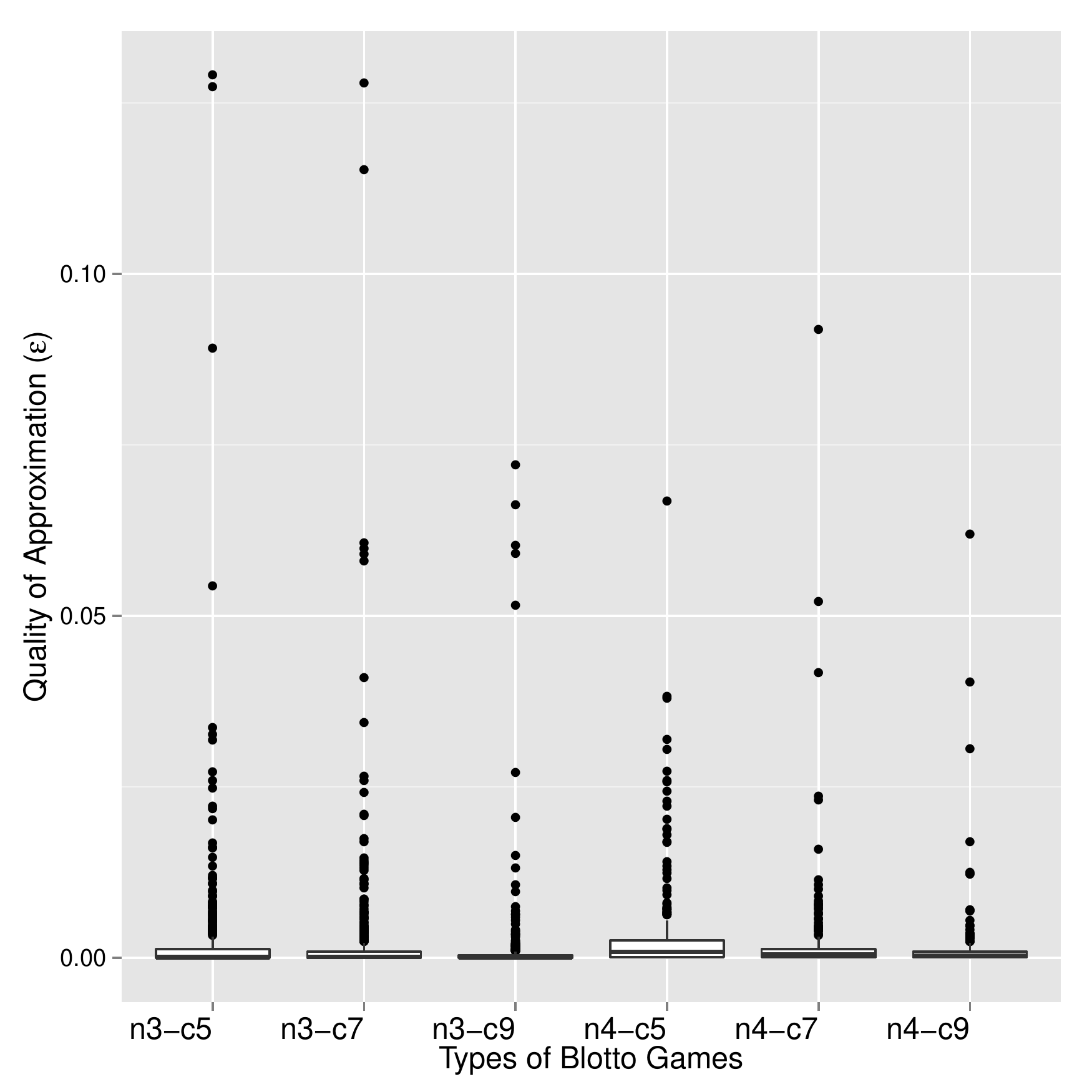}
\caption{Box and whisker plots for quality of $\epsilon$-NE found by \TSone on Blotto
games.}
\label{fig:box_b}
\end{figure}

The best theoretical upper bound on the performance of the TS algorithm is that
it produces a $0.3393$-Nash equilibrium, but previous work has not been able to
show a matching lower bound. As we have seen, the algorithm usually finds a very
good approximation in practice. The Blotto games were
the only class that challenged the algorithm, and even then
the approximations were usually good. Figure~\ref{fig:box_b} shows
box and whisker plots for the quality of approximation of \TSone on the 
Blotto games that we considered. Almost all points were
close to 0.01, and only a handful of instances were larger. The worst
approximation that we found was a $0.14$-Nash equilibrium, which is still far
from the worst-case guarantee.



To test the limits of the TS algorithm, we used a genetic algorithm to try and
find worst-case examples. More precisely, we used a genome that encoded a $5
\times 5$ game, and a fitness function that measured the quality of
approximation found by \TSone on the game. The result of this was a $5 \times 5$
game\footnote{All games found are given in the full version: 
\texttt{http://arxiv.org/abs/1502.04980}} for which \TSone produces a \worstepsts-Nash equilibrium.  Note that this
essentially matches the theoretical upper bound.

Although this example provides a good lower bound against \TSone, we found that
\BBMtwo produced a 0.024-NE when applied to this instance. For this reason, we
decided to test the limits of our entire portfolio of algorithms. We used the
same genetic algorithm, but this time the fitness function was the best of the
approximations found by \TSone, \Pure, and \BBMtwo. We produced a $5 \times 5$
game for which all of the above
algorithms produced at best a \worstepsmix-Nash equilibrium (TS produced a
\worstepsmix-NE, \Pure produced a 0.324-NE, and \BBMtwo produced a 0.321-NE.) 
 
Both of the games mentioned so far contain dominated strategies. While this does
not invalidate the lower bounds, it is obviously undesirable. For this reason,
we reran the experiments with a fitness function that penalizes dominated
strategies. For TS, we found a game with no dominated strategies for which the
algorithm produces a \worstndts-NE, and for the portfolio we found a game with
no dominated strategies for which the portfolio finds a \worstndport-NE.

Finally, we applied the genetic algorithm to try and find a worst case example
for \KSplus, but we were unsuccessful. We were able to produce a $5 \times 5$ game for
which \KSplus finds a \worstepsksplus-WSNE, but this is not particularly useful,
as some of the techniques used in this algorithm cannot possibly find an
$\epsilon$-WSNE where $\epsilon$ is better $0.5$.

\section{Conclusion}

In this paper, we have conducted experiments to test the applicability of
approximation methods in practice. We found that the existing library of games
provided by GAMUT is biased towards games that always have pure Nash equilibria,
so we introduced several new classes of games where this is not the case. Having
done so, we are able to give conclusive answers to the questions that we posed
in the introduction. Firstly, we have seen that the exact algorithms \LH and \SE
are quite limited in their ability to solve large games, particularly on the
games that we have introduced. Secondly, in contrast to this, we have seen that
approximation methods can tackle much larger instances, and that they provide 
approximate equilibria that are good enough to be practically useful. Finally, we have
seen that \DMP, BBM, and the variations of TS, can all be useful depending on
the quality of approximation that is required, which shows that there is a
trade-off between running time and quality of approximation. In addition to
this, we have also applied genetic algorithms to find a new worst case-example
for TS which essentially match the theoretical upper bound.

This work has highlighted the need for a comprehensive library of games upon
which game theoretic algorithms can be tested. In addition to the games that we
have introduced, there are also a number of other areas that could be
represented here. For example, there are many auction problems from
which games could be derived.  Our study has focussed on algorithms with
provable guarantees on the quality of approximate equilibria found in polynomial
time.  One direction for further study would to be to consider exact algorithms
as heuristics for finding approximate equilibria.  For example, one could
randomly sample supports and find the best approximate equilibria on these
supports using linear programming, or one could run an algorithm like \LH for a
fixed time or number of steps and check how good the strategy profiles it traces
are as approximate equilibria.  It would be interesting to see the extent that
genetic algorithms can be applied to this. For example, what happens when we try
to make \LH bad, for quality of approximation or for running time? For
approximation results, we would be interested in the approximation that is found
after the algorithm has taken a fixed number of steps, like linear, quadratic,
or some other polynomial.

{\small
\bibliographystyle{abbrv}
\bibliography{biblio}
}

\newpage
\appendix




\section{Appendix}

\subsection{Performance of TS}
\label{app:box}

\begin{figure}[ht]
\includegraphics[scale=0.6]{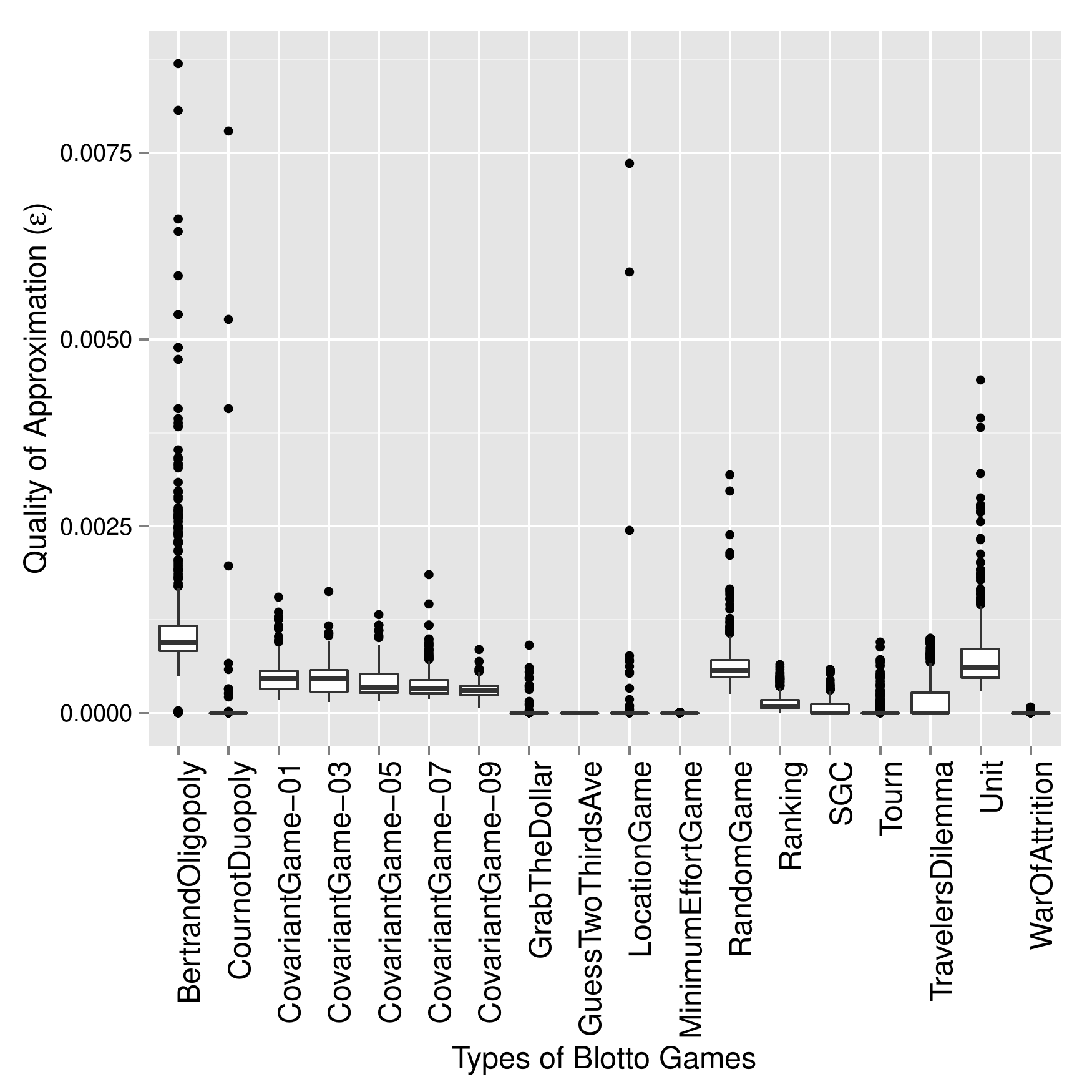}
\caption{Box and whisker plots for the quality of $\epsilon$-NE found by \TSone.}
\label{fig:box_g}
\end{figure}

\newpage

\subsection{Close to worst case examples for TS}

\begin{figure}[ht]
\centering
\begin{subfigure}[t]{0.49\textwidth}
\[
A =
\begin{bmatrix} 
56887 & 96209 & 18255 & 39791 & 34582\\
85196 & 65524 & 99782 & 97070 & 96765\\
72792 & 22633 & 59840 & 99422 & 20117\\
32963 & 18315 & 72881 & 85009 & 88426\\
26072 & 57677 & 19770 & 67548 & 43422\\
\end{bmatrix}
\]
\[
B =
\begin{bmatrix}
97209 & 43186 & 16342 & 16048 & 39059\\
60251 & 99413 & 3320 & 4596 & 28038\\
39941 & 78448 & 52995 & 10725 & 76914\\
91891 & 83644 & 53468 & 15871 & 16222\\
83130 & 99012 & 95399 & 95610 & 44184\\
\end{bmatrix}
\]
\caption{A bimatrix game for which \TSone finds a \worstepsts-NE.}
\label{fig:worstts}
\end{subfigure}
\begin{subfigure}[t]{0.49\textwidth}
\[
A =
\begin{bmatrix} 
60599 & 68002 & 94170 & 91622 & 92353 \\
25814 & 90811 & 37672 & 74870 & 45424 \\
45917 & 33775 & 23776 & 45343 & 71345\\
25339 & 61618 & 37098 & 24770 & 24292 \\
39656 & 24016 & 63463 & 63396 & 75634 \\
\end{bmatrix}
\]
\[
B =
\begin{bmatrix}
 59772 & 98580 & 12901 & 34113 & 34916\\
 98451 & 59776 & 20776 & 13893 & 27315\\
 17890 & 81316 & 36363 & 20947 & 36334\\
 53317 & 45329 & 14652 & 42939 & 17098\\
 16562 & 67590 & 62024 & 16783 & 13882\\
\end{bmatrix}
\]
\caption{A bimatrix game for which \TSone finds a 0.3189-NE, \Pure finds a
0.324-NE, and \BBMtwo finds a 0.321-NE.}
\label{fig:worstall}
\end{subfigure}
\begin{subfigure}[t]{0.49\textwidth}
\[
A= 
\begin{bmatrix}
98433 & 55567 & 47993 & 40394 & 29895\\
64392 & 81847 & 55756 & 40084 & 53038\\
83294 & 65160 & 57977 & 29881 & 85410\\
96761 & 64995 & 94650 & 66773 & 98497\\
49601 & 37356 & 53139 & 90306 & 40336\\
\end{bmatrix}
\]
\[
B= 
\begin{bmatrix}
23881 & 47704 & 29961 & 29770 & 30574\\
24025 & 29210 & 59730 & 62710 & 45525\\
37004 & 26562 & 87859 & 87746 & 49475\\
24815 & 25213 & 23425 & 94911 & 22725\\
89638 & 29302 & 95049 & 71163 & 41298\\
\end{bmatrix}
\]
\caption{A bimatrix game without dominated strategies which \TSone finds \worstndts-NE.}
\end{subfigure}
\begin{subfigure}[t]{0.49\textwidth}
\[
A= 
\begin{bmatrix}
98433 & 55567 & 47993 & 40394 & 29895\\
64392 & 81847 & 55756 & 40084 & 53038\\
83294 & 65160 & 57977 & 29881 & 85445\\
96761 & 64995 & 94650 & 66773 & 98497\\
49601 & 37356 & 53139 & 90306 & 40336\\
\end{bmatrix}
\]
\[
B= 
\begin{bmatrix}
23881 & 47704 & 29961 & 29770 & 30574\\
24025 & 29210 & 59730 & 62710 & 45525\\
37004 & 26562 & 87859 & 87746 & 49475\\
24815 & 25213 & 23425 & 94911 & 22725\\
89638 & 29302 & 95049 & 71163 & 41298\\
\end{bmatrix}
\]
\caption{A bimatrix game without dominated strategies which \TSone finds a 0.325-NE, \Pure finds a
0.329-NE and \BBMtwo finds a 0.374-NE.}
\end{subfigure}
\end{figure}

\end{document}